\documentclass[aps,prl,twocolumn,showpacs,superscriptaddress,floatfix]{revtex4-1}
\usepackage{amsmath}
\usepackage{graphicx}
\usepackage{epstopdf}
\usepackage{dcolumn}
\usepackage{soul}
\usepackage[version=4]{mhchem}
\usepackage{bm}
\usepackage{hyperref}
\usepackage{xcolor}
\usepackage{mathtools}
\usepackage{textcomp} 

\pdfoutput=1

\newcommand{\unit}{$\,$}
 
\newcommand{\dd}{\mathrm{d}}

\DeclareUnicodeCharacter{2212}{-}

\begin{document}

\title{Tunneling-tip-induced collapse of the charge gap in the excitonic insulator \ce{Ta$_2$NiSe$_5$}}

\author{Qingyu He}
\email{q.he@fkf.mpg.de}
\affiliation{Max Planck Institute for Solid State Research, 70569 Stuttgart, Germany}
\author{Xinglu Que}
\affiliation{Max Planck Institute for Solid State Research, 70569 Stuttgart, Germany}
\author{Lihui Zhou}

\affiliation{Max Planck Institute for Solid State Research, 70569 Stuttgart, Germany}
\author{Masahiko Isobe}
\affiliation{Max Planck Institute for Solid State Research, 70569 Stuttgart, Germany}
\author{Dennis Huang}
\affiliation{Max Planck Institute for Solid State Research, 70569 Stuttgart, Germany}
\author{Hidenori Takagi}
\email{h.takagi@fkf.mpg.de}
\affiliation{Max Planck Institute for Solid State Research, 70569 Stuttgart, Germany}
\affiliation{Department of Physics, University of Tokyo, 113-0033 Tokyo, Japan}
\affiliation{Institute for Functional Matter and Quantum Technologies, University of Stuttgart, 70569 Stuttgart, Germany}

\date{\today}
\begin{abstract}

Tuning many-body electronic phases by an external handle is of both fundamental and practical importance in condensed matter science. The tunability mirrors the underlying interactions, and gigantic electric, optical and magnetic responses to minute external stimuli can be anticipated in the critical region of phase change. The excitonic insulator is one of the more exotic phases of interacting electrons, produced by the Coulomb attraction between a small and equal number of electrons and holes, leading to the spontaneous formation of exciton pairs in narrow-gap semiconductors/semimetals. The layered chalcogenide \ce{Ta$_2$NiSe$_5$} has been recently discussed as a candidate for the excitonic insulator, though the nature of the excitation gap that opens below $T_c=328$\unit K remains hotly debated. Here, we demonstrate a drastic collapse of the excitation gap in \ce{Ta$_2$NiSe$_5$} and the realization of a zero-gap state by moving the tip of a cryogenic scanning tunneling microscope towards the sample surface by a few angstroms. We argue that the collapse of the gap is driven predominantly by the electrostatic charge accumulation at the surface induced by the proximity of the tip and the resultant carrier doping. The fragility of the gap in the insulating state of \ce{Ta$_2$NiSe$_5$} strongly suggests the gap possesses many-body character. Our results establish a novel, reversible phase-change function based on \ce{Ta$_2$NiSe$_5$}.

\end{abstract}
\maketitle{}

\begin{figure}
    \includegraphics[width=\columnwidth]{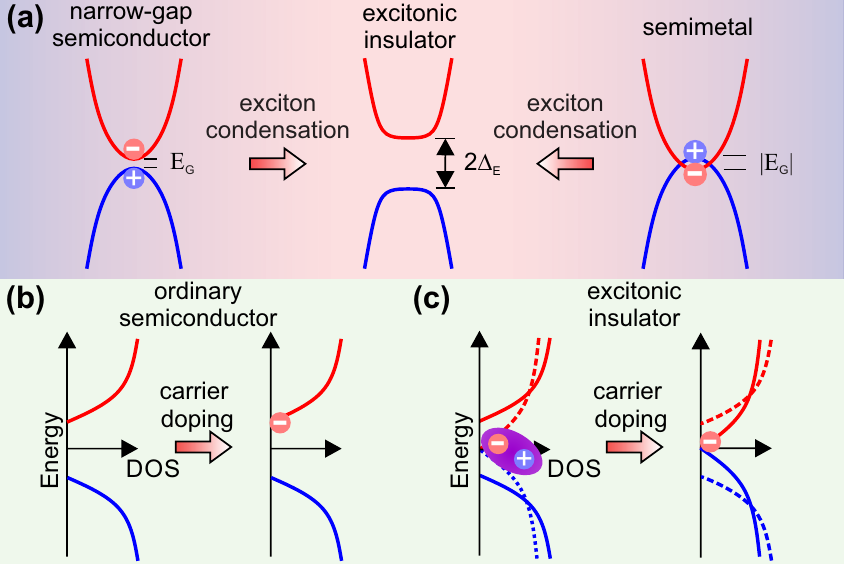}
    \caption{(a) Conceptual image of the transition from a narrow-gap semiconductor (left) or semimetal (right) to an excitonic insulator (middle). A many-body gap $2\Delta_E$ opens up in the excitonic phase. (b) and (c) Doping of an ordinary semiconductor or an excitonic insulator leading to a rigid shift of the chemical potential or the collapse of the energy gap, respectively.}
    \label{fig:1}
\end{figure}

First proposed roughly 60 years ago, the excitonic insulator is an exotic phase of interacting electrons expected to arise in narrow-gap semiconductors and semimetals~\cite{Mott1961,Kohn1967,Jerome1967}. In narrow-gap semiconductors, if the binding energy $E_B$ of electron-hole excitations due to the Coulomb attraction exceeds the single-electron band gap $E_G$, then electron-hole pairs spontaneously condense below a transition temperature $T_c$. The state below $T_c$ is called an excitonic insulator and hosts elementary excitations characterized by a many-body gap $2\Delta_E$, as conceptually illustrated in Fig. \ref{fig:1}(a). When $E_G$ is approximately zero, $T_c$ and $2\Delta_E$ are maximized, and when $E_G$ increases up to $E_B$, $T_c$ and $2\Delta_E$ decrease to zero. Even in semimetals with $E_G<0$, the excitonic transition can still take place, but $T_c$ and $2\Delta_E$ are suppressed with increasing band overlap $|E_G|$, since the attractive Coulomb interactions are rapidly screened out by the increasing number of holes and electrons. As a result, $T_c$ exhibits a dome-shaped dependence on $E_G$, with maximum at zero gap, $E_G=0$.

\begin{figure}
    \includegraphics[width=\columnwidth]{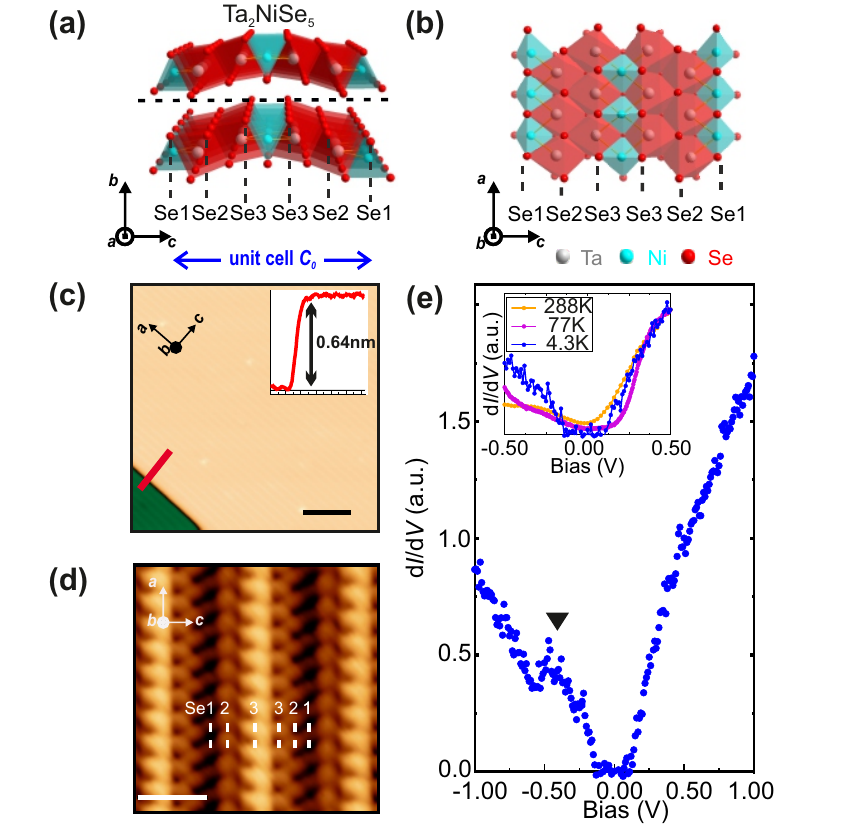}
    \caption{(a) and (b) Crystal structure of \ce{Ta$_2$NiSe$_5$} presented in two different views. The horizontal dashed line marks the cleavage plane. (c) Large-area topography of \ce{Ta$_2$NiSe$_5$} (53\unit nm $\times$ 53\unit nm, $I_s=100$\unit pA, $V_s=400$\unit mV, $T=77$\unit K). The insets show the correspondence to the crystallographic axes and the line-cut height profile through the step edge denoted by the red line. The black bar represents a length of 10\unit nm. (d) Small-area topography of \ce{Ta$_2$NiSe$_5$} with atomic resolution (3.5\unit nm $\times$ 3.5\unit nm, $I_s=1$\unit nA, $V_s=400$\unit mV, $T=77$\unit K). The insets show the correspondence to the crystal structure and crystallographic axes. The white bar represents a length of 1\unit nm. (e) Spectroscopic data taken at 4.3\unit K, exhibiting a charge gap and a terrace-like structure at negative biases, marked by the arrow. The inset displays the spectra taken at various temperatures, tracking the gap evolution.}
    \label{fig:2}
\end{figure}

Recent intensive studies have established the ternary layered chalcogenide \ce{Ta$_2$NiSe$_5$} as one of the most promising candidates for an excitonic insulator in a bulk material~\cite{Wakisaka2009,Kaneko2013,Kim2016,Lu2017,Larkin2017,Mor2017,Seo2018,
Werdehauseneaap2018,Mor2018,Okazaki2018,Nakano2018,Lee2019,Fukutani2019,Chen2020,Volkov2020,Kim2020,Mazza2020}. \ce{Ta$_2$NiSe$_5$} is electronically quasi-one-dimensional, as each layer consists of an alternating array of one chain of \ce{NiSe$_4$} tetrahedra and two chains of \ce{TaSe$_6$} octahedra [Figs. \ref{fig:2}(a) and \ref{fig:2}(b)]. It has a nearly zero or negative direct gap at the $\Gamma$ point \cite{Fukutani2019,Chen2020,Lu2017}, with \ce{Ta} 5$\it{d}$ orbitals comprising the conduction bands and \ce{Ni} 3$\it{d}$ (\ce{Se} 4$\it{p}$) orbitals comprising the valence bands. Below $T_c=328$ \unit K, \ce{Ta$_2$NiSe$_5$} undergoes a phase transition to an insulator state with a finite excitation gap~\cite{Lu2017,Seo2018}. Two initial observations supported the excitonic insulator nature of this gap. First, optical conductivity data revealed the formation of a charge gap~\cite{Lu2017} with an optical gap edge of 0.16\unit eV and an isosbestic point of gap opening $\sim$0.3\unit eV below $T_c$. The energy scale of this gap agreed reasonably well with the exciton binding energy $E_B\sim$ 0.25\unit eV estimated for the sister compound \ce{Ta$_2$NiS$_5$}, which is a band insulator with a large single-electron energy gap of 0.6\unit eV~\cite{Larkin2017}. Second, upon application of pressure or sulfur substitution, $E_G$ could be controlled, and $T_c$ traced out a dome peaked at $E_G\sim 0$, corresponding to pure \ce{Ta$_2$NiSe$_5$} at ambient pressure~\cite{Lu2017}. 

Nevertheless, \ce{Ta$_2$NiSe$_5$} undergoes a structural transition from the orthorhombic to the monoclinic phase at $T_c=328$ \unit K, without superlattice formation, concomitant with the electronic phase transition~\cite{Nakano2018,Salvo1986,Sunshine1985}. The hybridization of the \ce{Ta} conduction bands and the \ce{Ni} valence bands, which is forbidden by symmetry in the orthorhombic phase, is permitted in the monoclinic phase, giving rise to the possible opening of a one-electron gap. Recent symmetry analysis has shown that a finite excitonic order parameter must be accompanied by the monoclinic distortion~\cite{Mazza2020}. This means that the hybridization gap, which is single electron in nature, is inherently and inevitably involved in the formation of the excitation gap~\cite{Mazza2020,Watson2020,Subedi2020,Baldini2020}. Theoretical calculations for the monoclinic phase of \ce{Ta$_2$NiSe$_5$} predict a hybridization gap of up to 140 meV \cite{Watson2020,Subedi2020}, clearly falling short of the experimental value of the charge gap. Therefore, the structural phase transition alone cannot fully explain the excitation gap, and its excitonic character deserves more attention.

Further hints may come from examining the fragility of the insulating phase. As a general example, Mott insulators derived from repulsive Coulomb interactions between electrons often collapse into a metallic phase with a small number of carrier doping and/or application of pressure, reflecting the many-body character of their charge gap. Such control of electronic phases provides the basis for correlated electronics~\cite{Takagi2010,Ahn2003,Ahn2006,Ye2012,Lu2015,Lu2017,Deng2018}. The same concepts may be extended to many-body electronic systems with attractive interactions, namely, an excitonic insulator.  Several optical pumping studies have been devoted to the tuning of the insulating phase~\cite{Mor2018,Okazaki2018} in \ce{Ta$_2$NiSe$_5$} and an electronic phase evolution in the time domain has been observed. Angle-resolved photoemission spectroscopy (ARPES) measurements have also probed the effect of depositing \ce{K} atoms on the surface of \ce{Ta$_2$NiSe$_5$}, resulting in a Stark effect, carrier doping, or both~\cite{Fukutani2019,Chen2020}. As excitonic pairs consist of equal numbers of electrons and holes, we naturally expect that modest electron or hole doping would disturb the balance and thereby rapidly suppress the excitonic phase and the excitation gap.  In contrast, if the gap were predominantly a single-electron hybridization gap, as in a conventional semiconductor, the doping effect would not modify the gap, but only shift the chemical potential, as shown in Fig.~\ref{fig:1}(b). Thus, more experiments along this line, perhaps by tuning the carrier density in a local, reversible and clean manner, may illuminate whether exciton formation or hybridization due to the monoclinic distortion plays the primary role.

In this work, we utilized cryogenic scanning tunneling microscopy (STM) to trigger and to probe the phase change in the excitonic insulator candidate \ce{Ta$_2$NiSe$_5$}. We discovered an abrupt collapse of the excitation gap from $\sim$ 0.25\unit eV to zero upon moving the STM tip towards the sample surface by merely a few angstroms. We argue that the phase change is driven by tip-induced electrostatic surface charges, which generate an imbalance of electron and hole densities and cause the system to revert to a zero-gap semiconductor, as schematically shown in Fig. \ref{fig:1}(c). Our results strongly suggest that the gap in \ce{Ta$_2$NiSe$_5$} has many-body and excitonic character and demonstrate nanoscale, reversible phase-tuning in the compound. 

Single crystals of  \ce{Ta$_2$NiSe$_5$} were grown via the chemical vapour transport technique performed in a sealed quartz tube with \ce{I$_2$} as the transport agent, as described in Ref.~\cite{Salvo1986}. The transport and optical conductivity measurements clearly indicate the transition from a narrow-gap semiconductor/semimetal to an insulator at $T_c=328$\unit K, which we ascribed to an excitonic transition in our previous communications~\cite{Lu2017}.
All experiments were conducted in a home-built ultrahigh vacuum (UHV) STM facility with a base temperature of 4.3\unit K. The mounted crystal was cleaved at a low temperature ($\sim$ 200\unit K) in the UHV chamber, then quickly transferred to the microscope head held at 4.3\unit K. The topographic images were acquired in constant-current mode. The differential conductance $\dd I/\dd V$ was recorded using a standard lock-in technique, by adding a modulation voltage $V_m$ to a fixed bias, $V_s$. All data were collected at 4.3\unit K unless otherwise specified. 

The topographic STM image is fully consistent with the crystal structure determined for bulk single crystals [Figs. \ref{fig:2}(a) and \ref{fig:2}(b)] and indicates a high quality of surface for spectroscopic investigations. In the large-area image in Fig. \ref{fig:2}(c), we observe atomically flat and clean terraces with a defect density less than 0.002\unit nm$^{-2}$ ($\sim$0.001 per unit-cell in the topmost layer) and an atomic step with a height of $\sim$0.6\unit nm, which agrees well with the interlayer distance, a half of the out-of-plane lattice constant $b$ [Fig. \ref{fig:2}(c)]. The cleaved surface should represent \ce{Se} atoms located at the top of layer.  In the expanded image in Fig. \ref{fig:2}(d), we can clearly see \ce{Se} atoms at the expected positions and a rippling of the layer. Five \ce{Se} chains within one period of the rippling modulation can be identified: the two in-phase atomic chains on the ridge are assigned to \ce{Se}3, the two neighboring chains are assigned to \ce{Se}2 and their respective neighboring chains at the bottom of valley are assigned to \ce{Se}1. Though not visible, the \ce{Ni} chains below the surface \ce{Se} atoms should run along the \ce{Se}1 chain and between the two \ce{Se}3 chains, while the \ce{Ta} chains below should lie between the \ce{Se}1 and \ce{Se}2 chains and between the \ce{Se}2 and \ce{Se}3 chains. 

Measurements of the local density of states (LDOS) via $\dd I/\dd V$ spectroscopy on the well-defined surface reveal the emergence of a charge gap upon lowering the temperature below $T_c$. At 288 \unit K, barely below $T_c=328$ \unit K, only a V-shaped pseudogap feature with finite $\dd I/\dd V$ at the Fermi energy is observed (Fig. \ref{fig:2}(e) inset). The incomplete gap near $T_c$ can be ascribed to the combined effects of thermal and lifetime broadening, similar to those observed in superconductors. With further lowering of the temperature, however, a clear charge gap develops, and below 77\unit K, a nearly temperature-independent, full gap of $\sim$0.25\unit eV is manifested (Fig. \ref{fig:2}(e) inset). We note that all the LDOS spectra in Fig. \ref{fig:2} were taken at a relatively large tip-sample distances (estimated to be larger than 0.92 \unit nm), in order to minimize any effects from the tip. The spectra also did not depend appreciably on the lateral position of the tip in real space.  The LDOS spectra obtained in this work agree well with those reported in preceding STM studies~\cite{Kim2016,Lee2019}. The apparent shape difference is due to Feenstra normalization~\cite{SM}, and the gap of $\sim$ 0.25\unit eV is consistent with the optical gap (0.16\unit eV gap edge and 0.3\unit eV isosbestic point) reported in our previous work~\cite{Lu2017,Larkin2017}. In the valence band region below the gap (negative bias voltage), an additional terrace-like structure is superposed on the LDOS between roughly $-$0.2\unit eV to $-$0.5\unit eV, which also was observed in an earlier study and ascribed to impurities~\cite{Kim2016}. Our spectra are taken in a clean area distant from impurity atoms and the terrace structure is observed independent of the location of the tip.  As later discussed, we suggest that these features originate neither from band structure details nor impurity atoms, but from self-trapped states of tip-induced electrons.
\begin{figure}
    \includegraphics[width=\columnwidth]{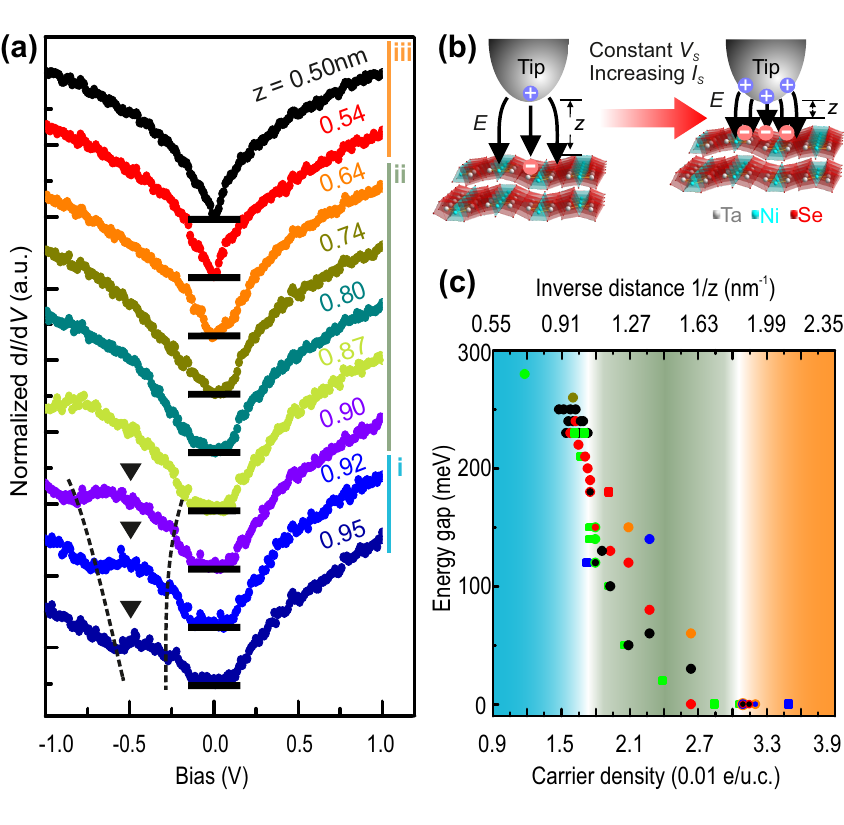}
    \caption{(a) The conductance $\dd I/\dd V$ spectra taken with various tip-sample distances $z$ from 0.95\unit nm to 0.50\unit nm. The $\dd I/\dd V$ curves are normalized by their maximum values and vertically shifted for clarity, with the horizontal black lines indicating the zero conductance for each curve. The terrace-like structure discussed in the text is marked by the black arrows and the dashed lines. Three regimes -- (i) stable 0.25\unit eV gap, (ii) gap shrinking and (iii) stable zero gap -- can be identified. (b) Schematic picture of the experimental setup. The tip-sample distance is controlled by varying the tunneling current $I_s$ while maintaining a fixed bias voltage $V_s$. The changing distance modifies the number of carriers induced by the difference in work function between the tip and the sample. (c) Experimentally determined charge gap as a function of the estimated surface carrier density and inverse distance. Different dataset are color- and symbol-coded: balls, $V_s=2$\unit V; squares, $V_s=1.5$\unit V.}
    \label{fig:3}
\end{figure}

Our main discovery is that the 0.25\unit eV charge gap in \ce{Ta$_2$NiSe$_5$} collapses rapidly upon moving the STM tip a few angstroms toward the surface and is therefore rather fragile against an external perturbation. The tip-sample distance $z$ in the STM setup is precisely controlled by varying the tunneling current $I_s$ while fixing the tip-sample voltage $V_s$ [as schematically shown in Fig. \ref{fig:3}(b)]. $I_s$ can be converted to a tip-sample distance $z$ following the procedure described in Ref. \cite{SM}. Fig. \ref{fig:3}(a) displays the $\dd I/\dd V$ spectra with various tip-sample distances $z$ and shows the collapse of the 0.25\unit eV gap upon lowering $z$ from 0.95 to 0.50 \unit nm.  When the tip is far from the surface ($\geq$0.92\unit nm), the $\dd I/\dd V$ spectra consistently show a $z$-independent gap of 0.25\unit eV and the perturbation by the tip does not influence the insulating ground state appreciably. In an intermediate $z$ range from 0.92\unit nm to 0.54\unit nm, however, the gap is apparently and very rapidly suppressed from 0.25\unit eV to almost zero. Eventually, with $z<0.54$\unit nm, the gap appears to collapse completely and a V-shaped LDOS, indicative of a zero-gap semiconducting state, emerges. The drastic change of LDOS upon reducing $z$ can be separated into those three regimes, as marked in Fig.~\ref{fig:3}(a).  This gap collapse is fully reversible~\cite{SM}. Another series of experiments also show the same collapse of the gap with decreasing $z$~\cite{SM}. We note that the tip-sample distances here $>0.5$\unit nm is large enough to eliminate the role of mechanical pressure by the tip in the observed gap collapse. Furthermore, mechanical contact would likely damage the tip or the sample surface, resulting in a loss of reversibility.

The approach of the tip towards the sample is accompanied by a systematic growth of the terrace-like feature observed around $-$0.2 to $-$0.5\unit eV in the $\dd I/\dd V$ spectra, starting at a large $z=0.95$\unit nm. This dramatic growth suggests that the terrace structure is not an intrinsic band structure feature of \ce{Ta$_2$NiSe$_5$}, but something else very likely connected to the mechanism of gap collapse.  While the gap does not change appreciably, the terrace spreads over a wider energy range as $z$ decreases to 0.92\unit nm. Below $z=0.92$\unit nm, the high-energy tail of terrace extends further and even into the gap region, which appears to trigger the collapse of gap.

We ascribe the observed collapse of the charge gap to a carrier-doping effect from tip-induced electrostatic charges at the sample surface~\cite{Jiang2017}. When the tip is brought in close proximity to the sample surface, an electric field is generated between the two due to the difference in their work functions. We use the literature value of the tip work function $\phi_T=4.50$\unit eV~\cite{Battisti2017} and extract the sample work function $\phi_S=5.05$ \unit eV from $I$-$z$ spectroscopy (Ref.~\cite{Ahn2003} estimated 5.4 eV), leading to a difference of 0.55 \unit eV. We argue that the resulting electric field traps electrons at the sample surface, which manifest as the terrace-like structure, 0.2--0.5\unit eV below the Fermi energy, in the $\dd I/\dd V$ spectrum at $z=0.95$\unit nm. The fact that the terrace appears only below the Fermi energy implies that electrons (not holes) are trapped at the sample surface, consistent with the sign of the work function difference. Furthermore, the energy scale of the terrace, which represents the binding energy of these trapped electrons, is comparable to the work function difference. The width of the terrace in $\dd I/\dd V$ then very likely corresponds to the energy dispersion of the trapped electrons. When $z$ decreases from 0.95\unit nm, the electric field and the number of trapped electrons increase, thereby causing the bandwidth of the terrace structure to grow [Fig. \ref{fig:3}(a)]. The terrace band eventually spreads into the bulk gap below $z=0.92$\unit nm and local carrier doping is expected, whereby the gap begins to collapse. At $z=0.92$\unit nm, the difference of work function gives an electric field of $6.1\times 10^6$\unit V/cm right below the tip and the electrostatically induced surface carrier density is estimated to be $3.4\times 10^{-2}$\unit e/nm$^2$, which corresponds $1.8\times 10^{-2}$\unit e/(unit cell) assuming that the trapped electrons are only at the topmost layer~\cite{SM}. This number is comparable to the population of electrons and holes participating in the formation of the (excitonic) charge gap~\cite{Lu2017}, but importantly, we note that only a fraction of the trapped surface electrons, those that lie in the high-energy tail of terrace structure that has extended into the gap region, contribute to doping. In other words, only a small number of effectively doped electrons, compared to the total number of electrons and holes involved, is needed to destabilize the charge gap. We note that the small number of effectively doped charges is consistent with the pinning of the chemical potential at the bottom of the valley, in between the conduction band and the valence band, in the zero-gap semiconductor region, $z<0.54$\unit nm.

Fig. \ref{fig:3}(c) summarizes the gap evolution in \ce{Ta$_2$NiSe$_5$} versus the tip distance and estimated surface charge density for several data sets, in both directions of increasing and decreasing $z$. We note that the drastic gap collapse is emphatically distinct from tip-induced band bending, which would result in a modest increase, not decrease, of the band gap size~\cite{SM, Marczinowski2008}. Our results are in principle consistent with \ce{K}-dosing ARPES experiments~\cite{Fukutani2019,Chen2020}, although the details differ, including the observation of a Stark effect in the latter, perhaps due to the different mechanisms of carrier doping. In our case, the instability of the 0.25\unit eV excitation gap to an STM tip points to it having excitonic insulating character. Only a small amount of carriers is necessary to generate an imbalance between electrons and holes and increase the screening of attractive Coulomb interactions, thereby collapsing the gap. We note that the gap is fully closed for z $<$ 0.54 \unit nm, with no remnant hybridization gap. As the monoclinic transition may be driven by the excitonic instability \cite{Kaneko2013}, it is possible that our tip-induced electronic transition drives a simultaneous structural transition, but we could not distinguish a loss of tiny monoclinic distortion in our STM images.

In conclusion, in our spectroscopic STM measurements of the putative excitonic insulator \ce{Ta$_2$NiSe$_5$}, we discovered a collapse of the 0.25\unit eV excitation gap in the LDOS upon moving the STM tip towards the surface by only a few tenth of nanometers. We argue that the collapse represents a local phase change from an insulator to a nearly zero-gap semiconductor, induced by local carrier doping due to a tip-induced electric field. The fragility of the 0.25\unit eV gap to the addition of a small amount of carriers strongly suggests it cannot be a pure single-electron band gap, but must have many-body, excitonic character. Our results also demonstrate a nanoscale, noninvasive, and reversible control of the charge gap in \ce{Ta$_2$NiSe$_5$}, which may be detected electronically and optically. This effect may form the basis of novel nanoscale phase-change devices with memory and sensor functions, wherein the insulating gap, not the carrier density, can be tuned, analogous to the case for a Mott transistor with a fragile many-body gap.

\begin{acknowledgements}
The authors thank A. Rost, T. Takayama, A. Yaresko, U. Wedig, M. Daghofer, and S. Loth for discussion and comments, and K. Pflaum and M. Dueller for technical assistance. This work has been supported in part by the Alexander von Humboldt foundation.
\end{acknowledgements}


\bibliography{TNS_STM.bib}

\begin{thebibliography}{35}%
\makeatletter
\providecommand \@ifxundefined [1]{%
 \@ifx{#1\undefined}
}%
\providecommand \@ifnum [1]{%
 \ifnum #1\expandafter \@firstoftwo
 \else \expandafter \@secondoftwo
 \fi
}%
\providecommand \@ifx [1]{%
 \ifx #1\expandafter \@firstoftwo
 \else \expandafter \@secondoftwo
 \fi
}%
\providecommand \natexlab [1]{#1}%
\providecommand \enquote  [1]{``#1''}%
\providecommand \bibnamefont  [1]{#1}%
\providecommand \bibfnamefont [1]{#1}%
\providecommand \citenamefont [1]{#1}%
\providecommand \href@noop [0]{\@secondoftwo}%
\providecommand \href [0]{\begingroup \@sanitize@url \@href}%
\providecommand \@href[1]{\@@startlink{#1}\@@href}%
\providecommand \@@href[1]{\endgroup#1\@@endlink}%
\providecommand \@sanitize@url [0]{\catcode `\\12\catcode `\$12\catcode
  `\&12\catcode `\#12\catcode `\^12\catcode `\_12\catcode `\%12\relax}%
\providecommand \@@startlink[1]{}%
\providecommand \@@endlink[0]{}%
\providecommand \url  [0]{\begingroup\@sanitize@url \@url }%
\providecommand \@url [1]{\endgroup\@href {#1}{\urlprefix }}%
\providecommand \urlprefix  [0]{URL }%
\providecommand \Eprint [0]{\href }%
\providecommand \doibase [0]{http://dx.doi.org/}%
\providecommand \selectlanguage [0]{\@gobble}%
\providecommand \bibinfo  [0]{\@secondoftwo}%
\providecommand \bibfield  [0]{\@secondoftwo}%
\providecommand \translation [1]{[#1]}%
\providecommand \BibitemOpen [0]{}%
\providecommand \bibitemStop [0]{}%
\providecommand \bibitemNoStop [0]{.\EOS\space}%
\providecommand \EOS [0]{\spacefactor3000\relax}%
\providecommand \BibitemShut  [1]{\csname bibitem#1\endcsname}%
\let\auto@bib@innerbib\@empty
\bibitem [{\citenamefont {Mott}(1961)}]{Mott1961}%
  \BibitemOpen
  \bibfield  {author} {\bibinfo {author} {\bibfnamefont {N.~F.}\ \bibnamefont
  {Mott}},\ }\href {\doibase 10.1080/14786436108243318} {\bibfield  {journal}
  {\bibinfo  {journal} {Philos. Mag.}\ }\textbf {\bibinfo {volume} {6}},\
  \bibinfo {pages} {287} (\bibinfo {year} {1961})}\BibitemShut {NoStop}%
\bibitem [{\citenamefont {Kohn}(1967)}]{Kohn1967}%
  \BibitemOpen
  \bibfield  {author} {\bibinfo {author} {\bibfnamefont {W.}~\bibnamefont
  {Kohn}},\ }\href {\doibase 10.1103/PhysRevLett.19.439} {\bibfield  {journal}
  {\bibinfo  {journal} {Phys. Rev. Lett.}\ }\textbf {\bibinfo {volume} {19}},\
  \bibinfo {pages} {439} (\bibinfo {year} {1967})}\BibitemShut {NoStop}%
\bibitem [{\citenamefont {J\'erome}\ \emph {et~al.}(1967)\citenamefont
  {J\'erome}, \citenamefont {Rice},\ and\ \citenamefont {Kohn}}]{Jerome1967}%
  \BibitemOpen
  \bibfield  {author} {\bibinfo {author} {\bibfnamefont {D.}~\bibnamefont
  {J\'erome}}, \bibinfo {author} {\bibfnamefont {T.~M.}\ \bibnamefont {Rice}},
  \ and\ \bibinfo {author} {\bibfnamefont {W.}~\bibnamefont {Kohn}},\ }\href
  {\doibase 10.1103/PhysRev.158.462} {\bibfield  {journal} {\bibinfo  {journal}
  {Phys. Rev.}\ }\textbf {\bibinfo {volume} {158}},\ \bibinfo {pages} {462}
  (\bibinfo {year} {1967})}\BibitemShut {NoStop}%
\bibitem [{\citenamefont {Wakisaka}\ \emph {et~al.}(2009)\citenamefont
  {Wakisaka}, \citenamefont {Sudayama}, \citenamefont {Takubo}, \citenamefont
  {Mizokawa}, \citenamefont {Arita}, \citenamefont {Namatame}, \citenamefont
  {Taniguchi}, \citenamefont {Katayama}, \citenamefont {Nohara},\ and\
  \citenamefont {Takagi}}]{Wakisaka2009}%
  \BibitemOpen
  \bibfield  {author} {\bibinfo {author} {\bibfnamefont {Y.}~\bibnamefont
  {Wakisaka}}, \bibinfo {author} {\bibfnamefont {T.}~\bibnamefont {Sudayama}},
  \bibinfo {author} {\bibfnamefont {K.}~\bibnamefont {Takubo}}, \bibinfo
  {author} {\bibfnamefont {T.}~\bibnamefont {Mizokawa}}, \bibinfo {author}
  {\bibfnamefont {M.}~\bibnamefont {Arita}}, \bibinfo {author} {\bibfnamefont
  {H.}~\bibnamefont {Namatame}}, \bibinfo {author} {\bibfnamefont
  {M.}~\bibnamefont {Taniguchi}}, \bibinfo {author} {\bibfnamefont
  {N.}~\bibnamefont {Katayama}}, \bibinfo {author} {\bibfnamefont
  {M.}~\bibnamefont {Nohara}}, \ and\ \bibinfo {author} {\bibfnamefont
  {H.}~\bibnamefont {Takagi}},\ }\href {\doibase
  10.1103/PhysRevLett.103.026402} {\bibfield  {journal} {\bibinfo  {journal}
  {Phys. Rev. Lett.}\ }\textbf {\bibinfo {volume} {103}},\ \bibinfo {pages}
  {026402} (\bibinfo {year} {2009})}\BibitemShut {NoStop}%
\bibitem [{\citenamefont {Kaneko}\ \emph {et~al.}(2013)\citenamefont {Kaneko},
  \citenamefont {Toriyama}, \citenamefont {Konishi},\ and\ \citenamefont
  {Ohta}}]{Kaneko2013}%
  \BibitemOpen
  \bibfield  {author} {\bibinfo {author} {\bibfnamefont {T.}~\bibnamefont
  {Kaneko}}, \bibinfo {author} {\bibfnamefont {T.}~\bibnamefont {Toriyama}},
  \bibinfo {author} {\bibfnamefont {T.}~\bibnamefont {Konishi}}, \ and\
  \bibinfo {author} {\bibfnamefont {Y.}~\bibnamefont {Ohta}},\ }\href {\doibase
  10.1103/PhysRevB.87.035121} {\bibfield  {journal} {\bibinfo  {journal} {Phys.
  Rev. B}\ }\textbf {\bibinfo {volume} {87}},\ \bibinfo {pages} {035121}
  (\bibinfo {year} {2013})}\BibitemShut {NoStop}%
\bibitem [{\citenamefont {Kim}\ \emph {et~al.}(2016)\citenamefont {Kim},
  \citenamefont {Kim}, \citenamefont {Kang}, \citenamefont {An}, \citenamefont
  {Kim}, \citenamefont {Eom}, \citenamefont {Lee}, \citenamefont {Park},
  \citenamefont {Kim}, \citenamefont {Choi}, \citenamefont {Min},\ and\
  \citenamefont {Kim}}]{Kim2016}%
  \BibitemOpen
  \bibfield  {author} {\bibinfo {author} {\bibfnamefont {S.~Y.}\ \bibnamefont
  {Kim}}, \bibinfo {author} {\bibfnamefont {Y.}~\bibnamefont {Kim}}, \bibinfo
  {author} {\bibfnamefont {C.-J.}\ \bibnamefont {Kang}}, \bibinfo {author}
  {\bibfnamefont {E.-S.}\ \bibnamefont {An}}, \bibinfo {author} {\bibfnamefont
  {H.~K.}\ \bibnamefont {Kim}}, \bibinfo {author} {\bibfnamefont {M.~J.}\
  \bibnamefont {Eom}}, \bibinfo {author} {\bibfnamefont {M.}~\bibnamefont
  {Lee}}, \bibinfo {author} {\bibfnamefont {C.}~\bibnamefont {Park}}, \bibinfo
  {author} {\bibfnamefont {T.-H.}\ \bibnamefont {Kim}}, \bibinfo {author}
  {\bibfnamefont {H.~C.}\ \bibnamefont {Choi}}, \bibinfo {author}
  {\bibfnamefont {B.~I.}\ \bibnamefont {Min}}, \ and\ \bibinfo {author}
  {\bibfnamefont {J.~S.}\ \bibnamefont {Kim}},\ }\href {\doibase
  10.1021/acsnano.6b04796} {\bibfield  {journal} {\bibinfo  {journal} {ACS
  Nano}\ }\textbf {\bibinfo {volume} {10}},\ \bibinfo {pages} {8888} (\bibinfo
  {year} {2016})}\BibitemShut {NoStop}%
\bibitem [{\citenamefont {Lu}\ \emph {et~al.}(2017)\citenamefont {Lu},
  \citenamefont {Kono}, \citenamefont {Larkin}, \citenamefont {Rost},
  \citenamefont {Takayama}, \citenamefont {Boris}, \citenamefont {Keimer},\
  and\ \citenamefont {Takagi}}]{Lu2017}%
  \BibitemOpen
  \bibfield  {author} {\bibinfo {author} {\bibfnamefont {Y.~F.}\ \bibnamefont
  {Lu}}, \bibinfo {author} {\bibfnamefont {H.}~\bibnamefont {Kono}}, \bibinfo
  {author} {\bibfnamefont {T.~I.}\ \bibnamefont {Larkin}}, \bibinfo {author}
  {\bibfnamefont {A.~W.}\ \bibnamefont {Rost}}, \bibinfo {author}
  {\bibfnamefont {T.}~\bibnamefont {Takayama}}, \bibinfo {author}
  {\bibfnamefont {A.~V.}\ \bibnamefont {Boris}}, \bibinfo {author}
  {\bibfnamefont {B.}~\bibnamefont {Keimer}}, \ and\ \bibinfo {author}
  {\bibfnamefont {H.}~\bibnamefont {Takagi}},\ }\href {\doibase
  10.1038/ncomms14408} {\bibfield  {journal} {\bibinfo  {journal} {Nat.
  Commun.}\ }\textbf {\bibinfo {volume} {8}},\ \bibinfo {pages} {14408}
  (\bibinfo {year} {2017})}\BibitemShut {NoStop}%
\bibitem [{\citenamefont {Larkin}\ \emph {et~al.}(2017)\citenamefont {Larkin},
  \citenamefont {Yaresko}, \citenamefont {Pr\"opper}, \citenamefont {Kikoin},
  \citenamefont {Lu}, \citenamefont {Takayama}, \citenamefont {Mathis},
  \citenamefont {Rost}, \citenamefont {Takagi}, \citenamefont {Keimer},\ and\
  \citenamefont {Boris}}]{Larkin2017}%
  \BibitemOpen
  \bibfield  {author} {\bibinfo {author} {\bibfnamefont {T.~I.}\ \bibnamefont
  {Larkin}}, \bibinfo {author} {\bibfnamefont {A.~N.}\ \bibnamefont {Yaresko}},
  \bibinfo {author} {\bibfnamefont {D.}~\bibnamefont {Pr\"opper}}, \bibinfo
  {author} {\bibfnamefont {K.~A.}\ \bibnamefont {Kikoin}}, \bibinfo {author}
  {\bibfnamefont {Y.~F.}\ \bibnamefont {Lu}}, \bibinfo {author} {\bibfnamefont
  {T.}~\bibnamefont {Takayama}}, \bibinfo {author} {\bibfnamefont {Y.-L.}\
  \bibnamefont {Mathis}}, \bibinfo {author} {\bibfnamefont {A.~W.}\
  \bibnamefont {Rost}}, \bibinfo {author} {\bibfnamefont {H.}~\bibnamefont
  {Takagi}}, \bibinfo {author} {\bibfnamefont {B.}~\bibnamefont {Keimer}}, \
  and\ \bibinfo {author} {\bibfnamefont {A.~V.}\ \bibnamefont {Boris}},\ }\href
  {\doibase 10.1103/PhysRevB.95.195144} {\bibfield  {journal} {\bibinfo
  {journal} {Phys. Rev. B}\ }\textbf {\bibinfo {volume} {95}},\ \bibinfo
  {pages} {195144} (\bibinfo {year} {2017})}\BibitemShut {NoStop}%
\bibitem [{\citenamefont {Mor}\ \emph {et~al.}(2017)\citenamefont {Mor},
  \citenamefont {Herzog}, \citenamefont {Gole\ifmmode~\check{z}\else
  \v{z}\fi{}}, \citenamefont {Werner}, \citenamefont {Eckstein}, \citenamefont
  {Katayama}, \citenamefont {Nohara}, \citenamefont {Takagi}, \citenamefont
  {Mizokawa}, \citenamefont {Monney},\ and\ \citenamefont
  {St\"ahler}}]{Mor2017}%
  \BibitemOpen
  \bibfield  {author} {\bibinfo {author} {\bibfnamefont {S.}~\bibnamefont
  {Mor}}, \bibinfo {author} {\bibfnamefont {M.}~\bibnamefont {Herzog}},
  \bibinfo {author} {\bibfnamefont {D.}~\bibnamefont
  {Gole\ifmmode~\check{z}\else \v{z}\fi{}}}, \bibinfo {author} {\bibfnamefont
  {P.}~\bibnamefont {Werner}}, \bibinfo {author} {\bibfnamefont
  {M.}~\bibnamefont {Eckstein}}, \bibinfo {author} {\bibfnamefont
  {N.}~\bibnamefont {Katayama}}, \bibinfo {author} {\bibfnamefont
  {M.}~\bibnamefont {Nohara}}, \bibinfo {author} {\bibfnamefont
  {H.}~\bibnamefont {Takagi}}, \bibinfo {author} {\bibfnamefont
  {T.}~\bibnamefont {Mizokawa}}, \bibinfo {author} {\bibfnamefont
  {C.}~\bibnamefont {Monney}}, \ and\ \bibinfo {author} {\bibfnamefont
  {J.}~\bibnamefont {St\"ahler}},\ }\href {\doibase
  10.1103/PhysRevLett.119.086401} {\bibfield  {journal} {\bibinfo  {journal}
  {Phys. Rev. Lett.}\ }\textbf {\bibinfo {volume} {119}},\ \bibinfo {pages}
  {086401} (\bibinfo {year} {2017})}\BibitemShut {NoStop}%
\bibitem [{\citenamefont {Seo}\ \emph {et~al.}(2018)\citenamefont {Seo},
  \citenamefont {Eom}, \citenamefont {Kim}, \citenamefont {Kang}, \citenamefont
  {Min},\ and\ \citenamefont {Hwang}}]{Seo2018}%
  \BibitemOpen
  \bibfield  {author} {\bibinfo {author} {\bibfnamefont {Y.-S.}\ \bibnamefont
  {Seo}}, \bibinfo {author} {\bibfnamefont {M.~J.}\ \bibnamefont {Eom}},
  \bibinfo {author} {\bibfnamefont {J.~S.}\ \bibnamefont {Kim}}, \bibinfo
  {author} {\bibfnamefont {C.-J.}\ \bibnamefont {Kang}}, \bibinfo {author}
  {\bibfnamefont {B.~I.}\ \bibnamefont {Min}}, \ and\ \bibinfo {author}
  {\bibfnamefont {J.}~\bibnamefont {Hwang}},\ }\href {\doibase
  10.1038/s41598-018-30430-9} {\bibfield  {journal} {\bibinfo  {journal} {Sci.
  Rep.}\ }\textbf {\bibinfo {volume} {8}},\ \bibinfo {pages} {11961} (\bibinfo
  {year} {2018})}\BibitemShut {NoStop}%
\bibitem [{\citenamefont {Werdehausen}\ \emph {et~al.}(2018)\citenamefont
  {Werdehausen}, \citenamefont {Takayama}, \citenamefont {H{\"o}ppner},
  \citenamefont {Albrecht}, \citenamefont {Rost}, \citenamefont {Lu},
  \citenamefont {Manske}, \citenamefont {Takagi},\ and\ \citenamefont
  {Kaiser}}]{Werdehauseneaap2018}%
  \BibitemOpen
  \bibfield  {author} {\bibinfo {author} {\bibfnamefont {D.}~\bibnamefont
  {Werdehausen}}, \bibinfo {author} {\bibfnamefont {T.}~\bibnamefont
  {Takayama}}, \bibinfo {author} {\bibfnamefont {M.}~\bibnamefont
  {H{\"o}ppner}}, \bibinfo {author} {\bibfnamefont {G.}~\bibnamefont
  {Albrecht}}, \bibinfo {author} {\bibfnamefont {A.~W.}\ \bibnamefont {Rost}},
  \bibinfo {author} {\bibfnamefont {Y.}~\bibnamefont {Lu}}, \bibinfo {author}
  {\bibfnamefont {D.}~\bibnamefont {Manske}}, \bibinfo {author} {\bibfnamefont
  {H.}~\bibnamefont {Takagi}}, \ and\ \bibinfo {author} {\bibfnamefont
  {S.}~\bibnamefont {Kaiser}},\ }\href {\doibase 10.1126/sciadv.aap8652}
  {\bibfield  {journal} {\bibinfo  {journal} {Sci. Adv.}\ }\textbf {\bibinfo
  {volume} {4}} (\bibinfo {year} {2018}),\ 10.1126/sciadv.aap8652}\BibitemShut
  {NoStop}%
\bibitem [{\citenamefont {Mor}\ \emph {et~al.}(2018)\citenamefont {Mor},
  \citenamefont {Herzog}, \citenamefont {Noack}, \citenamefont {Katayama},
  \citenamefont {Nohara}, \citenamefont {Takagi}, \citenamefont {Trunschke},
  \citenamefont {Mizokawa}, \citenamefont {Monney},\ and\ \citenamefont
  {St\"ahler}}]{Mor2018}%
  \BibitemOpen
  \bibfield  {author} {\bibinfo {author} {\bibfnamefont {S.}~\bibnamefont
  {Mor}}, \bibinfo {author} {\bibfnamefont {M.}~\bibnamefont {Herzog}},
  \bibinfo {author} {\bibfnamefont {J.}~\bibnamefont {Noack}}, \bibinfo
  {author} {\bibfnamefont {N.}~\bibnamefont {Katayama}}, \bibinfo {author}
  {\bibfnamefont {M.}~\bibnamefont {Nohara}}, \bibinfo {author} {\bibfnamefont
  {H.}~\bibnamefont {Takagi}}, \bibinfo {author} {\bibfnamefont
  {A.}~\bibnamefont {Trunschke}}, \bibinfo {author} {\bibfnamefont
  {T.}~\bibnamefont {Mizokawa}}, \bibinfo {author} {\bibfnamefont
  {C.}~\bibnamefont {Monney}}, \ and\ \bibinfo {author} {\bibfnamefont
  {J.}~\bibnamefont {St\"ahler}},\ }\href {\doibase 10.1103/PhysRevB.97.115154}
  {\bibfield  {journal} {\bibinfo  {journal} {Phys. Rev. B}\ }\textbf {\bibinfo
  {volume} {97}},\ \bibinfo {pages} {115154} (\bibinfo {year}
  {2018})}\BibitemShut {NoStop}%
\bibitem [{\citenamefont {Okazaki}\ \emph {et~al.}(2018)\citenamefont
  {Okazaki}, \citenamefont {Ogawa}, \citenamefont {Suzuki}, \citenamefont
  {Yamamoto}, \citenamefont {Someya}, \citenamefont {Michimae}, \citenamefont
  {Watanabe}, \citenamefont {Lu}, \citenamefont {Nohara}, \citenamefont
  {Takagi}, \citenamefont {Katayama}, \citenamefont {Sawa}, \citenamefont
  {Fujisawa}, \citenamefont {Kanai}, \citenamefont {Ishii}, \citenamefont
  {Itatani}, \citenamefont {Mizokawa},\ and\ \citenamefont
  {Shin}}]{Okazaki2018}%
  \BibitemOpen
  \bibfield  {author} {\bibinfo {author} {\bibfnamefont {K.}~\bibnamefont
  {Okazaki}}, \bibinfo {author} {\bibfnamefont {Y.}~\bibnamefont {Ogawa}},
  \bibinfo {author} {\bibfnamefont {T.}~\bibnamefont {Suzuki}}, \bibinfo
  {author} {\bibfnamefont {T.}~\bibnamefont {Yamamoto}}, \bibinfo {author}
  {\bibfnamefont {T.}~\bibnamefont {Someya}}, \bibinfo {author} {\bibfnamefont
  {S.}~\bibnamefont {Michimae}}, \bibinfo {author} {\bibfnamefont
  {M.}~\bibnamefont {Watanabe}}, \bibinfo {author} {\bibfnamefont
  {Y.}~\bibnamefont {Lu}}, \bibinfo {author} {\bibfnamefont {M.}~\bibnamefont
  {Nohara}}, \bibinfo {author} {\bibfnamefont {H.}~\bibnamefont {Takagi}},
  \bibinfo {author} {\bibfnamefont {N.}~\bibnamefont {Katayama}}, \bibinfo
  {author} {\bibfnamefont {H.}~\bibnamefont {Sawa}}, \bibinfo {author}
  {\bibfnamefont {M.}~\bibnamefont {Fujisawa}}, \bibinfo {author}
  {\bibfnamefont {T.}~\bibnamefont {Kanai}}, \bibinfo {author} {\bibfnamefont
  {N.}~\bibnamefont {Ishii}}, \bibinfo {author} {\bibfnamefont
  {J.}~\bibnamefont {Itatani}}, \bibinfo {author} {\bibfnamefont
  {T.}~\bibnamefont {Mizokawa}}, \ and\ \bibinfo {author} {\bibfnamefont
  {S.}~\bibnamefont {Shin}},\ }\href {\doibase 10.1038/s41467-018-06801-1}
  {\bibfield  {journal} {\bibinfo  {journal} {Nat. Commun.}\ }\textbf {\bibinfo
  {volume} {9}},\ \bibinfo {pages} {4322} (\bibinfo {year} {2018})}\BibitemShut
  {NoStop}%
\bibitem [{\citenamefont {Nakano}\ \emph {et~al.}(2018)\citenamefont {Nakano},
  \citenamefont {Hasegawa}, \citenamefont {Tamura}, \citenamefont {Katayama},
  \citenamefont {Tsutsui},\ and\ \citenamefont {Sawa}}]{Nakano2018}%
  \BibitemOpen
  \bibfield  {author} {\bibinfo {author} {\bibfnamefont {A.}~\bibnamefont
  {Nakano}}, \bibinfo {author} {\bibfnamefont {T.}~\bibnamefont {Hasegawa}},
  \bibinfo {author} {\bibfnamefont {S.}~\bibnamefont {Tamura}}, \bibinfo
  {author} {\bibfnamefont {N.}~\bibnamefont {Katayama}}, \bibinfo {author}
  {\bibfnamefont {S.}~\bibnamefont {Tsutsui}}, \ and\ \bibinfo {author}
  {\bibfnamefont {H.}~\bibnamefont {Sawa}},\ }\href {\doibase
  10.1103/PhysRevB.98.045139} {\bibfield  {journal} {\bibinfo  {journal} {Phys.
  Rev. B}\ }\textbf {\bibinfo {volume} {98}},\ \bibinfo {pages} {045139}
  (\bibinfo {year} {2018})}\BibitemShut {NoStop}%
\bibitem [{\citenamefont {Lee}\ \emph {et~al.}(2019)\citenamefont {Lee},
  \citenamefont {Kang}, \citenamefont {Eom}, \citenamefont {Kim}, \citenamefont
  {Min},\ and\ \citenamefont {Yeom}}]{Lee2019}%
  \BibitemOpen
  \bibfield  {author} {\bibinfo {author} {\bibfnamefont {J.}~\bibnamefont
  {Lee}}, \bibinfo {author} {\bibfnamefont {C.-J.}\ \bibnamefont {Kang}},
  \bibinfo {author} {\bibfnamefont {M.~J.}\ \bibnamefont {Eom}}, \bibinfo
  {author} {\bibfnamefont {J.~S.}\ \bibnamefont {Kim}}, \bibinfo {author}
  {\bibfnamefont {B.~I.}\ \bibnamefont {Min}}, \ and\ \bibinfo {author}
  {\bibfnamefont {H.~W.}\ \bibnamefont {Yeom}},\ }\href {\doibase
  10.1103/PhysRevB.99.075408} {\bibfield  {journal} {\bibinfo  {journal} {Phys.
  Rev. B}\ }\textbf {\bibinfo {volume} {99}},\ \bibinfo {pages} {075408}
  (\bibinfo {year} {2019})}\BibitemShut {NoStop}%
\bibitem [{\citenamefont {Fukutani}\ \emph {et~al.}(2019)\citenamefont
  {Fukutani}, \citenamefont {Stania}, \citenamefont {Jung}, \citenamefont
  {Schwier}, \citenamefont {Shimada}, \citenamefont {Kwon}, \citenamefont
  {Kim},\ and\ \citenamefont {Yeom}}]{Fukutani2019}%
  \BibitemOpen
  \bibfield  {author} {\bibinfo {author} {\bibfnamefont {K.}~\bibnamefont
  {Fukutani}}, \bibinfo {author} {\bibfnamefont {R.}~\bibnamefont {Stania}},
  \bibinfo {author} {\bibfnamefont {J.}~\bibnamefont {Jung}}, \bibinfo {author}
  {\bibfnamefont {E.~F.}\ \bibnamefont {Schwier}}, \bibinfo {author}
  {\bibfnamefont {K.}~\bibnamefont {Shimada}}, \bibinfo {author} {\bibfnamefont
  {C.~I.}\ \bibnamefont {Kwon}}, \bibinfo {author} {\bibfnamefont {J.~S.}\
  \bibnamefont {Kim}}, \ and\ \bibinfo {author} {\bibfnamefont {H.~W.}\
  \bibnamefont {Yeom}},\ }\href {\doibase 10.1103/PhysRevLett.123.206401}
  {\bibfield  {journal} {\bibinfo  {journal} {Phys. Rev. Lett.}\ }\textbf
  {\bibinfo {volume} {123}},\ \bibinfo {pages} {206401} (\bibinfo {year}
  {2019})}\BibitemShut {NoStop}%
\bibitem [{\citenamefont {Chen}\ \emph {et~al.}(2020)\citenamefont {Chen},
  \citenamefont {Han}, \citenamefont {Cai}, \citenamefont {Wang}, \citenamefont
  {Wang}, \citenamefont {Xin},\ and\ \citenamefont {Zhang}}]{Chen2020}%
  \BibitemOpen
  \bibfield  {author} {\bibinfo {author} {\bibfnamefont {L.}~\bibnamefont
  {Chen}}, \bibinfo {author} {\bibfnamefont {T.~T.}\ \bibnamefont {Han}},
  \bibinfo {author} {\bibfnamefont {C.}~\bibnamefont {Cai}}, \bibinfo {author}
  {\bibfnamefont {Z.~G.}\ \bibnamefont {Wang}}, \bibinfo {author}
  {\bibfnamefont {Y.~D.}\ \bibnamefont {Wang}}, \bibinfo {author}
  {\bibfnamefont {Z.~M.}\ \bibnamefont {Xin}}, \ and\ \bibinfo {author}
  {\bibfnamefont {Y.}~\bibnamefont {Zhang}},\ }\href {\doibase
  10.1103/PhysRevB.102.161116} {\bibfield  {journal} {\bibinfo  {journal}
  {Phys. Rev. B}\ }\textbf {\bibinfo {volume} {102}},\ \bibinfo {pages}
  {161116(R)} (\bibinfo {year} {2020})}\BibitemShut {NoStop}%
\bibitem [{\citenamefont {Volkov}\ \emph {et~al.}(2020)\citenamefont {Volkov},
  \citenamefont {Ye}, \citenamefont {Lohani}, \citenamefont {Feldman},
  \citenamefont {Kanigel}, \citenamefont {Haule},\ and\ \citenamefont
  {Blumberg}}]{Volkov2020}%
  \BibitemOpen
  \bibfield  {author} {\bibinfo {author} {\bibfnamefont {P.~A.}\ \bibnamefont
  {Volkov}}, \bibinfo {author} {\bibfnamefont {M.}~\bibnamefont {Ye}}, \bibinfo
  {author} {\bibfnamefont {H.}~\bibnamefont {Lohani}}, \bibinfo {author}
  {\bibfnamefont {I.}~\bibnamefont {Feldman}}, \bibinfo {author} {\bibfnamefont
  {A.}~\bibnamefont {Kanigel}}, \bibinfo {author} {\bibfnamefont
  {K.}~\bibnamefont {Haule}}, \ and\ \bibinfo {author} {\bibfnamefont
  {G.}~\bibnamefont {Blumberg}},\ }\href@noop {} {\  (\bibinfo {year}
  {2020})},\ \Eprint {http://arxiv.org/abs/2007.07344} {arXiv:2007.07344
  [cond-mat.str-el]} \BibitemShut {NoStop}%
\bibitem [{\citenamefont {Kim}\ \emph {et~al.}(2020)\citenamefont {Kim},
  \citenamefont {Schulz}, \citenamefont {Takayama}, \citenamefont {Isobe},
  \citenamefont {Takagi},\ and\ \citenamefont {Kaiser}}]{Kim2020}%
  \BibitemOpen
  \bibfield  {author} {\bibinfo {author} {\bibfnamefont {M.-J.}\ \bibnamefont
  {Kim}}, \bibinfo {author} {\bibfnamefont {A.}~\bibnamefont {Schulz}},
  \bibinfo {author} {\bibfnamefont {T.}~\bibnamefont {Takayama}}, \bibinfo
  {author} {\bibfnamefont {M.}~\bibnamefont {Isobe}}, \bibinfo {author}
  {\bibfnamefont {H.}~\bibnamefont {Takagi}}, \ and\ \bibinfo {author}
  {\bibfnamefont {S.}~\bibnamefont {Kaiser}},\ }\href {\doibase
  10.1103/PhysRevResearch.2.042039} {\bibfield  {journal} {\bibinfo  {journal}
  {Phys. Rev. Research}\ }\textbf {\bibinfo {volume} {2}},\ \bibinfo {pages}
  {042039(R)} (\bibinfo {year} {2020})}\BibitemShut {NoStop}%
\bibitem [{\citenamefont {Mazza}\ \emph {et~al.}(2020)\citenamefont {Mazza},
  \citenamefont {R\"osner}, \citenamefont {Windg\"atter}, \citenamefont
  {Latini}, \citenamefont {H\"ubener}, \citenamefont {Millis}, \citenamefont
  {Rubio},\ and\ \citenamefont {Georges}}]{Mazza2020}%
  \BibitemOpen
  \bibfield  {author} {\bibinfo {author} {\bibfnamefont {G.}~\bibnamefont
  {Mazza}}, \bibinfo {author} {\bibfnamefont {M.}~\bibnamefont {R\"osner}},
  \bibinfo {author} {\bibfnamefont {L.}~\bibnamefont {Windg\"atter}}, \bibinfo
  {author} {\bibfnamefont {S.}~\bibnamefont {Latini}}, \bibinfo {author}
  {\bibfnamefont {H.}~\bibnamefont {H\"ubener}}, \bibinfo {author}
  {\bibfnamefont {A.~J.}\ \bibnamefont {Millis}}, \bibinfo {author}
  {\bibfnamefont {A.}~\bibnamefont {Rubio}}, \ and\ \bibinfo {author}
  {\bibfnamefont {A.}~\bibnamefont {Georges}},\ }\href {\doibase
  10.1103/PhysRevLett.124.197601} {\bibfield  {journal} {\bibinfo  {journal}
  {Phys. Rev. Lett.}\ }\textbf {\bibinfo {volume} {124}},\ \bibinfo {pages}
  {197601} (\bibinfo {year} {2020})}\BibitemShut {NoStop}%
\bibitem [{\citenamefont {{Di Salvo}}\ \emph {et~al.}(1986)\citenamefont {{Di
  Salvo}}, \citenamefont {Chen}, \citenamefont {Fleming}, \citenamefont
  {Waszczak}, \citenamefont {Dunn}, \citenamefont {Sunshine},\ and\
  \citenamefont {Ibers}}]{Salvo1986}%
  \BibitemOpen
  \bibfield  {author} {\bibinfo {author} {\bibfnamefont {F.}~\bibnamefont {{Di
  Salvo}}}, \bibinfo {author} {\bibfnamefont {C.}~\bibnamefont {Chen}},
  \bibinfo {author} {\bibfnamefont {R.}~\bibnamefont {Fleming}}, \bibinfo
  {author} {\bibfnamefont {J.}~\bibnamefont {Waszczak}}, \bibinfo {author}
  {\bibfnamefont {R.}~\bibnamefont {Dunn}}, \bibinfo {author} {\bibfnamefont
  {S.}~\bibnamefont {Sunshine}}, \ and\ \bibinfo {author} {\bibfnamefont
  {J.~A.}\ \bibnamefont {Ibers}},\ }\href {\doibase
  https://doi.org/10.1016/0022-5088(86)90216-X} {\bibfield  {journal} {\bibinfo
   {journal} {J. Less Common Met.}\ }\textbf {\bibinfo {volume} {116}},\
  \bibinfo {pages} {51} (\bibinfo {year} {1986})}\BibitemShut {NoStop}%
\bibitem [{\citenamefont {Sunshine}\ and\ \citenamefont
  {Ibers}(1985)}]{Sunshine1985}%
  \BibitemOpen
  \bibfield  {author} {\bibinfo {author} {\bibfnamefont {S.~A.}\ \bibnamefont
  {Sunshine}}\ and\ \bibinfo {author} {\bibfnamefont {J.~A.}\ \bibnamefont
  {Ibers}},\ }\href {\doibase 10.1021/ic00216a027} {\bibfield  {journal}
  {\bibinfo  {journal} {Inorg. Chem.}\ }\textbf {\bibinfo {volume} {24}},\
  \bibinfo {pages} {3611} (\bibinfo {year} {1985})}\BibitemShut {NoStop}%
\bibitem [{\citenamefont {Watson}\ \emph {et~al.}(2020)\citenamefont {Watson},
  \citenamefont {Markovi\ifmmode~\acute{c}\else \'{c}\fi{}}, \citenamefont
  {Morales}, \citenamefont {Le~F\`evre}, \citenamefont {Merz}, \citenamefont
  {Haghighirad},\ and\ \citenamefont {King}}]{Watson2020}%
  \BibitemOpen
  \bibfield  {author} {\bibinfo {author} {\bibfnamefont {M.~D.}\ \bibnamefont
  {Watson}}, \bibinfo {author} {\bibfnamefont {I.}~\bibnamefont
  {Markovi\ifmmode~\acute{c}\else \'{c}\fi{}}}, \bibinfo {author}
  {\bibfnamefont {E.~A.}\ \bibnamefont {Morales}}, \bibinfo {author}
  {\bibfnamefont {P.}~\bibnamefont {Le~F\`evre}}, \bibinfo {author}
  {\bibfnamefont {M.}~\bibnamefont {Merz}}, \bibinfo {author} {\bibfnamefont
  {A.~A.}\ \bibnamefont {Haghighirad}}, \ and\ \bibinfo {author} {\bibfnamefont
  {P.~D.~C.}\ \bibnamefont {King}},\ }\href {\doibase
  10.1103/PhysRevResearch.2.013236} {\bibfield  {journal} {\bibinfo  {journal}
  {Phys. Rev. Research}\ }\textbf {\bibinfo {volume} {2}},\ \bibinfo {pages}
  {013236} (\bibinfo {year} {2020})}\BibitemShut {NoStop}%
\bibitem [{\citenamefont {Subedi}(2020)}]{Subedi2020}%
  \BibitemOpen
  \bibfield  {author} {\bibinfo {author} {\bibfnamefont {A.}~\bibnamefont
  {Subedi}},\ }\href {\doibase 10.1103/PhysRevMaterials.4.083601} {\bibfield
  {journal} {\bibinfo  {journal} {Phys. Rev. Materials}\ }\textbf {\bibinfo
  {volume} {4}},\ \bibinfo {pages} {083601} (\bibinfo {year}
  {2020})}\BibitemShut {NoStop}%
\bibitem [{\citenamefont {Baldini}\ \emph {et~al.}(2020)\citenamefont
  {Baldini}, \citenamefont {Zong}, \citenamefont {Choi}, \citenamefont {Lee},
  \citenamefont {Michael}, \citenamefont {Windgaetter}, \citenamefont {Mazin},
  \citenamefont {Latini}, \citenamefont {Azoury}, \citenamefont {Lv},
  \citenamefont {Kogar}, \citenamefont {Wang}, \citenamefont {Lu},
  \citenamefont {Takayama}, \citenamefont {Takagi}, \citenamefont {Millis},
  \citenamefont {Rubio}, \citenamefont {Demler},\ and\ \citenamefont
  {Gedik}}]{Baldini2020}%
  \BibitemOpen
  \bibfield  {author} {\bibinfo {author} {\bibfnamefont {E.}~\bibnamefont
  {Baldini}}, \bibinfo {author} {\bibfnamefont {A.}~\bibnamefont {Zong}},
  \bibinfo {author} {\bibfnamefont {D.}~\bibnamefont {Choi}}, \bibinfo {author}
  {\bibfnamefont {C.}~\bibnamefont {Lee}}, \bibinfo {author} {\bibfnamefont
  {M.~H.}\ \bibnamefont {Michael}}, \bibinfo {author} {\bibfnamefont
  {L.}~\bibnamefont {Windgaetter}}, \bibinfo {author} {\bibfnamefont {I.~I.}\
  \bibnamefont {Mazin}}, \bibinfo {author} {\bibfnamefont {S.}~\bibnamefont
  {Latini}}, \bibinfo {author} {\bibfnamefont {D.}~\bibnamefont {Azoury}},
  \bibinfo {author} {\bibfnamefont {B.}~\bibnamefont {Lv}}, \bibinfo {author}
  {\bibfnamefont {A.}~\bibnamefont {Kogar}}, \bibinfo {author} {\bibfnamefont
  {Y.}~\bibnamefont {Wang}}, \bibinfo {author} {\bibfnamefont {Y.}~\bibnamefont
  {Lu}}, \bibinfo {author} {\bibfnamefont {T.}~\bibnamefont {Takayama}},
  \bibinfo {author} {\bibfnamefont {H.}~\bibnamefont {Takagi}}, \bibinfo
  {author} {\bibfnamefont {A.~J.}\ \bibnamefont {Millis}}, \bibinfo {author}
  {\bibfnamefont {A.}~\bibnamefont {Rubio}}, \bibinfo {author} {\bibfnamefont
  {E.}~\bibnamefont {Demler}}, \ and\ \bibinfo {author} {\bibfnamefont
  {N.}~\bibnamefont {Gedik}},\ }\href@noop {} {\  (\bibinfo {year} {2020})},\
  \Eprint {http://arxiv.org/abs/2007.02909} {arXiv:2007.02909
  [cond-mat.str-el]} \BibitemShut {NoStop}%
\bibitem [{\citenamefont {Takagi}\ and\ \citenamefont
  {Hwang}(2010)}]{Takagi2010}%
  \BibitemOpen
  \bibfield  {author} {\bibinfo {author} {\bibfnamefont {H.}~\bibnamefont
  {Takagi}}\ and\ \bibinfo {author} {\bibfnamefont {H.~Y.}\ \bibnamefont
  {Hwang}},\ }\href {\doibase 10.1126/science.1228006} {\bibfield  {journal}
  {\bibinfo  {journal} {Science}\ }\textbf {\bibinfo {volume} {327}},\ \bibinfo
  {pages} {1601} (\bibinfo {year} {2010})}\BibitemShut {NoStop}%
\bibitem [{\citenamefont {Ahn}\ \emph {et~al.}(2003)\citenamefont {Ahn},
  \citenamefont {Triscone},\ and\ \citenamefont {Mannhart}}]{Ahn2003}%
  \BibitemOpen
  \bibfield  {author} {\bibinfo {author} {\bibfnamefont {C.~H.}\ \bibnamefont
  {Ahn}}, \bibinfo {author} {\bibfnamefont {J.-M.}\ \bibnamefont {Triscone}}, \
  and\ \bibinfo {author} {\bibfnamefont {J.}~\bibnamefont {Mannhart}},\ }\href
  {\doibase 10.1038/nature01878} {\bibfield  {journal} {\bibinfo  {journal}
  {Nature}\ }\textbf {\bibinfo {volume} {424}},\ \bibinfo {pages} {1015}
  (\bibinfo {year} {2003})}\BibitemShut {NoStop}%
\bibitem [{\citenamefont {Ahn}\ \emph {et~al.}(2006)\citenamefont {Ahn},
  \citenamefont {Bhattacharya}, \citenamefont {Di~Ventra}, \citenamefont
  {Eckstein}, \citenamefont {Frisbie}, \citenamefont {Gershenson},
  \citenamefont {Goldman}, \citenamefont {Inoue}, \citenamefont {Mannhart},
  \citenamefont {Millis}, \citenamefont {Morpurgo}, \citenamefont {Natelson},\
  and\ \citenamefont {Triscone}}]{Ahn2006}%
  \BibitemOpen
  \bibfield  {author} {\bibinfo {author} {\bibfnamefont {C.~H.}\ \bibnamefont
  {Ahn}}, \bibinfo {author} {\bibfnamefont {A.}~\bibnamefont {Bhattacharya}},
  \bibinfo {author} {\bibfnamefont {M.}~\bibnamefont {Di~Ventra}}, \bibinfo
  {author} {\bibfnamefont {J.~N.}\ \bibnamefont {Eckstein}}, \bibinfo {author}
  {\bibfnamefont {C.~D.}\ \bibnamefont {Frisbie}}, \bibinfo {author}
  {\bibfnamefont {M.~E.}\ \bibnamefont {Gershenson}}, \bibinfo {author}
  {\bibfnamefont {A.~M.}\ \bibnamefont {Goldman}}, \bibinfo {author}
  {\bibfnamefont {I.~H.}\ \bibnamefont {Inoue}}, \bibinfo {author}
  {\bibfnamefont {J.}~\bibnamefont {Mannhart}}, \bibinfo {author}
  {\bibfnamefont {A.~J.}\ \bibnamefont {Millis}}, \bibinfo {author}
  {\bibfnamefont {A.~F.}\ \bibnamefont {Morpurgo}}, \bibinfo {author}
  {\bibfnamefont {D.}~\bibnamefont {Natelson}}, \ and\ \bibinfo {author}
  {\bibfnamefont {J.-M.}\ \bibnamefont {Triscone}},\ }\href {\doibase
  10.1103/RevModPhys.78.1185} {\bibfield  {journal} {\bibinfo  {journal} {Rev.
  Mod. Phys.}\ }\textbf {\bibinfo {volume} {78}},\ \bibinfo {pages} {1185}
  (\bibinfo {year} {2006})}\BibitemShut {NoStop}%
\bibitem [{\citenamefont {Ye}\ \emph {et~al.}(2012)\citenamefont {Ye},
  \citenamefont {Zhang}, \citenamefont {Akashi}, \citenamefont {Bahramy},
  \citenamefont {Arita},\ and\ \citenamefont {Iwasa}}]{Ye2012}%
  \BibitemOpen
  \bibfield  {author} {\bibinfo {author} {\bibfnamefont {J.~T.}\ \bibnamefont
  {Ye}}, \bibinfo {author} {\bibfnamefont {Y.~J.}\ \bibnamefont {Zhang}},
  \bibinfo {author} {\bibfnamefont {R.}~\bibnamefont {Akashi}}, \bibinfo
  {author} {\bibfnamefont {M.~S.}\ \bibnamefont {Bahramy}}, \bibinfo {author}
  {\bibfnamefont {R.}~\bibnamefont {Arita}}, \ and\ \bibinfo {author}
  {\bibfnamefont {Y.}~\bibnamefont {Iwasa}},\ }\href {\doibase
  10.1126/science.1228006} {\bibfield  {journal} {\bibinfo  {journal}
  {Science}\ }\textbf {\bibinfo {volume} {338}},\ \bibinfo {pages} {1193}
  (\bibinfo {year} {2012})}\BibitemShut {NoStop}%
\bibitem [{\citenamefont {Lu}\ \emph {et~al.}(2015)\citenamefont {Lu},
  \citenamefont {Zheliuk}, \citenamefont {Leermakers}, \citenamefont {Yuan},
  \citenamefont {Zeitler}, \citenamefont {Law},\ and\ \citenamefont
  {Ye}}]{Lu2015}%
  \BibitemOpen
  \bibfield  {author} {\bibinfo {author} {\bibfnamefont {J.~M.}\ \bibnamefont
  {Lu}}, \bibinfo {author} {\bibfnamefont {O.}~\bibnamefont {Zheliuk}},
  \bibinfo {author} {\bibfnamefont {I.}~\bibnamefont {Leermakers}}, \bibinfo
  {author} {\bibfnamefont {N.~F.~Q.}\ \bibnamefont {Yuan}}, \bibinfo {author}
  {\bibfnamefont {U.}~\bibnamefont {Zeitler}}, \bibinfo {author} {\bibfnamefont
  {K.~T.}\ \bibnamefont {Law}}, \ and\ \bibinfo {author} {\bibfnamefont
  {J.~T.}\ \bibnamefont {Ye}},\ }\href {\doibase 10.1126/science.aab2277}
  {\bibfield  {journal} {\bibinfo  {journal} {Science}\ }\textbf {\bibinfo
  {volume} {350}},\ \bibinfo {pages} {1353} (\bibinfo {year}
  {2015})}\BibitemShut {NoStop}%
\bibitem [{\citenamefont {Deng}\ \emph {et~al.}(2018)\citenamefont {Deng},
  \citenamefont {Yu}, \citenamefont {Song}, \citenamefont {Zhang},
  \citenamefont {Wang}, \citenamefont {Sun}, \citenamefont {Yi}, \citenamefont
  {Wu}, \citenamefont {Wu}, \citenamefont {Zhu}, \citenamefont {Wang},
  \citenamefont {Chen},\ and\ \citenamefont {Zhang}}]{Deng2018}%
  \BibitemOpen
  \bibfield  {author} {\bibinfo {author} {\bibfnamefont {Y.}~\bibnamefont
  {Deng}}, \bibinfo {author} {\bibfnamefont {Y.}~\bibnamefont {Yu}}, \bibinfo
  {author} {\bibfnamefont {Y.}~\bibnamefont {Song}}, \bibinfo {author}
  {\bibfnamefont {J.}~\bibnamefont {Zhang}}, \bibinfo {author} {\bibfnamefont
  {N.~Z.}\ \bibnamefont {Wang}}, \bibinfo {author} {\bibfnamefont
  {Z.}~\bibnamefont {Sun}}, \bibinfo {author} {\bibfnamefont {Y.}~\bibnamefont
  {Yi}}, \bibinfo {author} {\bibfnamefont {Y.~Z.}\ \bibnamefont {Wu}}, \bibinfo
  {author} {\bibfnamefont {S.}~\bibnamefont {Wu}}, \bibinfo {author}
  {\bibfnamefont {J.}~\bibnamefont {Zhu}}, \bibinfo {author} {\bibfnamefont
  {J.}~\bibnamefont {Wang}}, \bibinfo {author} {\bibfnamefont {X.~H.}\
  \bibnamefont {Chen}}, \ and\ \bibinfo {author} {\bibfnamefont
  {Y.}~\bibnamefont {Zhang}},\ }\href {\doibase 10.1038/s41586-018-0626-9}
  {\bibfield  {journal} {\bibinfo  {journal} {Nature}\ }\textbf {\bibinfo
  {volume} {563}},\ \bibinfo {pages} {94} (\bibinfo {year} {2018})}\BibitemShut
  {NoStop}%
\bibitem [{SM()}]{SM}%
  \BibitemOpen
  \href@noop {} {\enquote {\bibinfo {title} {{See Supplemental Material at [URL
  will be inserted by publisher] for methods, additional data, and additional
  discussions.}}}\ }\BibitemShut {NoStop}%
\bibitem [{\citenamefont {Jiang}\ \emph {et~al.}(2017)\citenamefont {Jiang},
  \citenamefont {Mao}, \citenamefont {Moldovan}, \citenamefont {Masir},
  \citenamefont {Li}, \citenamefont {Watanabe}, \citenamefont {Taniguchi},
  \citenamefont {Peeters},\ and\ \citenamefont {Andrei}}]{Jiang2017}%
  \BibitemOpen
  \bibfield  {author} {\bibinfo {author} {\bibfnamefont {Y.}~\bibnamefont
  {Jiang}}, \bibinfo {author} {\bibfnamefont {J.}~\bibnamefont {Mao}}, \bibinfo
  {author} {\bibfnamefont {D.}~\bibnamefont {Moldovan}}, \bibinfo {author}
  {\bibfnamefont {M.~R.}\ \bibnamefont {Masir}}, \bibinfo {author}
  {\bibfnamefont {G.}~\bibnamefont {Li}}, \bibinfo {author} {\bibfnamefont
  {K.}~\bibnamefont {Watanabe}}, \bibinfo {author} {\bibfnamefont
  {T.}~\bibnamefont {Taniguchi}}, \bibinfo {author} {\bibfnamefont {F.~M.}\
  \bibnamefont {Peeters}}, \ and\ \bibinfo {author} {\bibfnamefont {E.~Y.}\
  \bibnamefont {Andrei}},\ }\href {\doibase 10.1038/nnano.2017.181} {\bibfield
  {journal} {\bibinfo  {journal} {Nat. Nanotechnol.}\ }\textbf {\bibinfo
  {volume} {12}},\ \bibinfo {pages} {1045} (\bibinfo {year}
  {2017})}\BibitemShut {NoStop}%
\bibitem [{\citenamefont {Battisti}\ \emph {et~al.}(2017)\citenamefont
  {Battisti}, \citenamefont {Fedoseev}, \citenamefont {Bastiaans},
  \citenamefont {de~la Torre}, \citenamefont {Perry}, \citenamefont
  {Baumberger},\ and\ \citenamefont {Allan}}]{Battisti2017}%
  \BibitemOpen
  \bibfield  {author} {\bibinfo {author} {\bibfnamefont {I.}~\bibnamefont
  {Battisti}}, \bibinfo {author} {\bibfnamefont {V.}~\bibnamefont {Fedoseev}},
  \bibinfo {author} {\bibfnamefont {K.~M.}\ \bibnamefont {Bastiaans}}, \bibinfo
  {author} {\bibfnamefont {A.}~\bibnamefont {de~la Torre}}, \bibinfo {author}
  {\bibfnamefont {R.~S.}\ \bibnamefont {Perry}}, \bibinfo {author}
  {\bibfnamefont {F.}~\bibnamefont {Baumberger}}, \ and\ \bibinfo {author}
  {\bibfnamefont {M.~P.}\ \bibnamefont {Allan}},\ }\href {\doibase
  10.1103/PhysRevB.95.235141} {\bibfield  {journal} {\bibinfo  {journal} {Phys.
  Rev. B}\ }\textbf {\bibinfo {volume} {95}},\ \bibinfo {pages} {235141}
  (\bibinfo {year} {2017})}\BibitemShut {NoStop}%
\bibitem [{\citenamefont {Marczinowski}\ \emph {et~al.}(2008)\citenamefont
  {Marczinowski}, \citenamefont {Wiebe}, \citenamefont {Meier}, \citenamefont
  {Hashimoto},\ and\ \citenamefont {Wiesendanger}}]{Marczinowski2008}%
  \BibitemOpen
  \bibfield  {author} {\bibinfo {author} {\bibfnamefont {F.}~\bibnamefont
  {Marczinowski}}, \bibinfo {author} {\bibfnamefont {J.}~\bibnamefont {Wiebe}},
  \bibinfo {author} {\bibfnamefont {F.}~\bibnamefont {Meier}}, \bibinfo
  {author} {\bibfnamefont {K.}~\bibnamefont {Hashimoto}}, \ and\ \bibinfo
  {author} {\bibfnamefont {R.}~\bibnamefont {Wiesendanger}},\ }\href {\doibase
  10.1103/PhysRevB.77.115318} {\bibfield  {journal} {\bibinfo  {journal} {Phys.
  Rev. B}\ }\textbf {\bibinfo {volume} {77}},\ \bibinfo {pages} {115318}
  (\bibinfo {year} {2008})}\BibitemShut {NoStop}%
\end{thebibliography}

\clearpage
\end{document}